\documentclass[journal=nalefd,manuscript=article]{achemso}

\usepackage{chemformula} % Formula subscripts using \ch{}
\usepackage[T1]{fontenc} % Use modern font encodings
\usepackage{graphicx}
\usepackage{siunitx}
\usepackage[utf8]{inputenc}
\usepackage[english]{babel}
\usepackage{amsmath}
\usepackage{natbib}
\usepackage{todonotes}
\usepackage{textcomp}

\author{Eva~Sch\"oll}
\altaffiliation{E.~Sch\"oll, L.~Hanschke, and L.~Schweickert contributed equally to this work.}
\affiliation{Department of Applied Physics, Royal Institute of Technology, Albanova University Centre, Roslagstullsbacken 21, 106 91 Stockholm, Sweden}
\author{Lukas~Hanschke}
\altaffiliation{E.~Sch\"oll, L.~Hanschke, and L.~Schweickert contributed equally to this work.}
\affiliation{Walter Schottky Institut and Physik Department, Technische Universit\"at München, 85748, Garching, Germany}
\author{Lucas~Schweickert}
\altaffiliation{E.~Sch\"oll, L.~Hanschke, and L.~Schweickert contributed equally to this work.}
\author{Katharina~D.~Zeuner}
\affiliation{Department of Applied Physics, Royal Institute of Technology, Albanova University Centre, Roslagstullsbacken 21, 106 91 Stockholm, Sweden}%
\author{Marcus~Reindl}
\author{Saimon~Filipe~Covre~da~Silva} 
\affiliation{Institute of Semiconductor and Solid State Physics, Johannes Kepler University Linz, 4040, Austria}
\author{Thomas~Lettner}
\affiliation{Department of Applied Physics, Royal Institute of Technology, Albanova University Centre, Roslagstullsbacken 21, 106 91 Stockholm, Sweden}%
\author{Rinaldo~Trotta}
\affiliation{Dipartimento di Fisica, Sapienza Universit\`a di Roma, Piazzale A. Moro 1, I-00185 Roma, Italy}
\author{Jonathan~J.~Finley}
\author{Kai~M\"uller}
\affiliation{Walter Schottky Institut and Physik Department, Technische Universit\"at München, 85748, Garching, Germany}
\author{Armando~Rastelli}
\affiliation{Institute of Semiconductor and Solid State Physics, Johannes Kepler University Linz, 4040, Austria}
\author{Val~Zwiller}
\author{Klaus~D.~J\"ons}
\email{klausj@kth.se}
\affiliation{Department of Applied Physics, Royal Institute of Technology, Albanova University Centre, Roslagstullsbacken 21, 106 91 Stockholm, Sweden}%

\title{Resonance fluorescence of GaAs quantum dots with near--unity photon indistinguishability}


\begin{document}

%\begin{tocentry}

%Some journals require a graphical entry for the Table of Contents.
%This should be laid out ``print ready'' so that the sizing of the

%Inside the \texttt{tocentry} environment, the font used is Helvetica
%8\,pt, as required by \emph{Journal of the American Chemical

%The surrounding frame is 9\,cm by 3.5\,cm, which is the maximum
%permitted for  \emph{Journal of the American Chemical Society}
%graphical table of content entries. The box will not resize if the
%content is too big: instead it will overflow the edge of the box.

%This box and the associated title will always be printed on a
%separate page at the end of the document.

%\end{tocentry}

\begin{abstract}
Photonic quantum technologies call for scalable quantum light sources that can be integrated, while providing the end user with single and entangled photons on--demand. One promising candidate are strain free GaAs/AlGaAs quantum dots obtained by droplet etching. Such quantum dots exhibit ultra low multi--photon probability and an unprecedented degree of photon pair entanglement. However, different to commonly studied InGaAs/GaAs quantum dots obtained by the Stranski-Krastanow mode, photons with a near--unity indistinguishability from these quantum emitters have proven to be elusive so far. Here, we show on--demand generation of near--unity indistinguishable photons from these quantum emitters by exploring pulsed resonance fluorescence. Given the short intrinsic lifetime of excitons confined in the GaAs quantum dots, we show single photon indistinguishability with a raw visibility of $V_{raw}=(94.2\pm5.2)\,\%$, without the need for Purcell enhancement.
Our results represent a milestone in the advance of GaAs quantum dots by demonstrating the final missing property standing in the way of using these emitters as a key component in quantum communication applications, e.g. as an entangled source for quantum repeater architectures.
\end{abstract}

Most applications in photonic quantum technologies rely on clean quantum interference of deterministically generated single and entangled photons. Quantum indistinghuishability is a crucial ingredient for the creation of higher N00N states~\cite{Walther.Pan.ea:2004,Nagata.Okamoto.ea:2007}, quantum teleportation~\cite{Bouwmeester.Pan.ea:1997} and swapping operations~\cite{Pan.Bouwmeester.ea:1998}, boson--sampling~\cite{Spring.Metcalf.ea:2013,Broome.Fedrizzi.ea:2013} and photon based quantum simulations~\cite{Aspuru-Guzik.Walther:2012}. An ideal quantum light source thus needs to emit photons on--demand, with high purity and near--unity indistinguishability as well as being scalable and interconnected with different quantum systems. Semiconductor quantum dots are proving to be the best sources that fulfill these requirements~\cite{Senellart.Solomon.ea:2017}, delivering ultra bright sources of on--demand single--photons at high rates compatible with photonic circuitry. Recently, GaAs quantum dots obtained by infilling of nanoholes produced by local droplet etching~\cite{Huber.Reindl.ea:2017} have emerged to be one of the most promising deterministic solid--state quantum light sources, reporting the lowest multi--photon probability~\cite{Schweickert.Joens.ea:2018}, the highest degree of polarization entanglement~\cite{Huber.Reindl.ea:2018} while also being the brightest entangled photon pair source reported~\cite{Chen.Zopf.ea:2018}. Entangled photons from these quantum dots have also been used to implement quantum teleportation protocols~\cite{Reindl.Huber.ea:2018}, thus proving their potential for quantum network applications. In particular, their emission wavelength range makes them suitable for hybrid quantum photonic technologies since they can be tuned into resonance with quantum memories, e.g. rubidium atoms~\cite{Akopian.Wang.ea:2011, 2018arXiv180805921S}. However, near--unity indistinguishable photons, a crucial element for photonic quantum technologies, was missing from this type of quantum dots until now.
Strong charge fluctuations at the vicinity of the quantum dot~\cite{Heyn.Zocher.ea:2017,Liu.Konthasinghe.ea:2018}, induced by the droplet growth method~\cite{Heyn.Stemmann.ea:2009} and faster phonon--induced pure dephasing~\cite{Axt.Kuhn.ea:2005} and zero--phonon broadening~\cite{Tighineanu.Dreessen.ea:2018} were suspected to be the main cause of the lower quantum interference visibility for these quantum emitters. In this letter, we apply cross--polarized pulsed resonance fluorescence to show that quantum dots derived from droplet etching do not suffer from additional dephasing mechanism at short time scales and exhibit near--unity indistinguishability of on--demand generated single photons. Remarkably, high quantum interference visibility values of $V_\text{raw}=(94.2\pm5.2)\,\%$ are achieved without the need of Purcell enhancement using microcavities~\cite{Somaschi.Giesz.ea:2016,Ding.He.ea:2016,Bennett.Lee.ea:2016}. Instead we capitalize on the intrinsically short lifetime of the excited states of these quantum emitters~\cite{Kiraz.Atatuere.ea:2004}.

\begin{figure}[htbp]
    \centering
    \includegraphics[width=170mm]{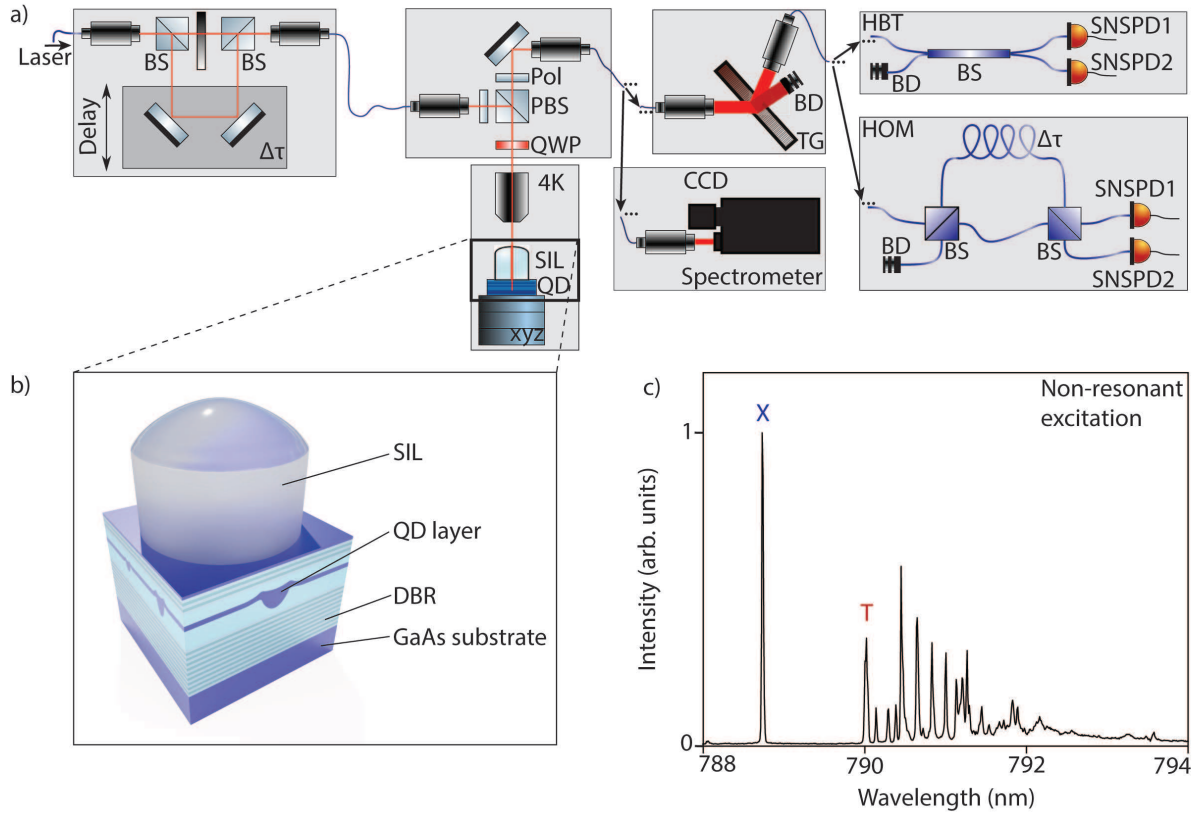}
    \caption{a) Modular setup consisting of laser excitation with delay line, confocal microscopy setup with polarization suppression, transmission spectrometer, Hanbury-Brown and Twiss setup (HBT) and Hong--Ou--Mandel setup (HOM). BS: beam splitter, BD: beam dump, TG: transmission grating, SNSPD: superconducting nanowire single photon detector, Pol: polarizer, PBS: polarizing beam splitter, QWP: quarter waveplate, HWP: half waveplate, SIL: solid immersion lens.  b) Schematic illustration of the sample structure. c) Spectrum of QD1 under non--resonant excitation.
    }
    \label{fig:figure1}
\end{figure}

%SETUP
For pulsed resonant s--shell excitation~\cite{He.He.ea:2013}, a polarization suppression setup was constructed similar to Ref~\cite{Kuhlmann.Houel.ea:2013}, but with the second polarizing beam splitter (PBS) replaced by a nanoparticle linear film polarizer, as illustrated in Fig.\,\ref{fig:figure1}~a). 
The sample was mounted in a low vibration closed--cycle cryostat and cooled to $5\,\si{\kelvin}$.
For excitation, a tunable, linear--polarized laser was used having a repetition rate of $80\,\si{\MHz}$ and pulse duration of $2.7\,\si{\pico\second}$. The excitation beam was directed onto the sample via the polarizing beam splitter and through an objective with $NA = 0.81$ and focused onto the quantum dot of interest using a solid immersion lens (SIL), directly attached to the sample surface. 
The signal was collected through the same optics in a confocal geometry and separated from the backscattered excitation laser by the polarizing beam splitter and a linear polarizer oriented perpendicular to the laser polarization. Further improvement of the laser suppression was achieved by spatial filtering since a small portion of the backscattered laser has a component perpendicular to the original polarization with a four--leaf clover patterned beam profile~\cite{Novotny:01}.

To perform photoluminescence measurements, the signal was coupled through a spectrometer onto a Silicon CCD.
For correlation measurements, the resonance fluorescence signal was further filtered with a home--built transmission spectrometer having a bandwidth of $19\,\si{\GHz}$ and an end--to--end efficiency exceeding $60\,\si{\percent}$.
Second--order intensity correlation measurements, were carried out with a Hanbury-Brown and Twiss type setup realized with a fiber coupled 50:50 beam splitter connected to two superconducting nanowire single photon detectors (SNSPD) with efficiencies of $50\,\si{\percent}$ and $60\,\si{\percent}$, a timing jitter of $20\,\si{\ps}$ and $30\,\si{\ps}$ and dark count rates of $0.006\,\text{dcts}/\si{\s}$ and $0.017\,\text{dcts}/\si{\s}$. 	
The detection events are recorded in a timetag file along with laser excitation events and analyzed with our Extensible Timetag Analyzer (ETA) software~\cite{ETA}.
To determine the indistinguishability of two consecutively emitted photons, two--photon interference is measured in a Hong--Ou--Mandel (HOM) type experiment. In order to interfere, these photons must impinge on a beam splitter with excellent spatial and temporal overlap. The temporal overlap is achieved by sending the signal into an unbalanced fiber--based Mach--Zehnder interferometer with a path--length difference of $2\,\si{\nano\second}$. The two output ports of the Mach--Zehnder interferometer are connected to a SNSPD each. Depending on the paths the photons take, they can arrive on the beam splitter simultaneously or with a time delay of $2\,\si{\ns}$ or $4\,\si{\ns}$, resulting in the characteristic quintuplet for Hong--Ou--Mandel measurements~\cite{Santori.Fattal.ea:2002} in the histogram.
The temporal overlap of the photons on the second beam splitter is ensured by splitting the excitation pulse into two identical pulses using another unbalanced Mach--Zehnder interferometer with variable delay. This delay is precisely tuned to the fixed fiber delay in the Hong--Ou--Mandel setup by measuring the interference of overlapping $1.94\,\si{\ps}$ short laser pulses using the same detectors as for all the correlation measurements. 

%SAMPLE
The GaAs quantum dot sample investigated in this work was grown by molecular beam epitaxy using the Aluminum droplet etching technique. The quantum dot layer was embedded in a $\lambda$--cavity made of $\text{Al}_{0.4}\text{Ga}_{0.6}\text{As}$ ($123\,\si{\nm}$) with 9.5 pairs of bottom and 2.5 pairs of top distributed Bragg reflectors (DBR) consisting of $\text{Al}_{0.95}\text{Ga}_{0.05}\text{As}$/$\text{Al}_{0.2}\text{Ga}_{0.8}\text{As}$ $\lambda/4$--thick layers as depicted in Fig\,\ref{fig:figure1}~b). The structure was finished by a $4\,\si{\nm}$ thick protective GaAs layer.
The Al--droplet etching method allows the growth of highly symmetric quantum dots with a low fine structure splitting (FSS)~\cite{Huo.Rastelli.ea:2013}.

%RESONANTSONANT EXCITATION

\begin{figure}[htbp]
    \centering
    \includegraphics[width=170mm]{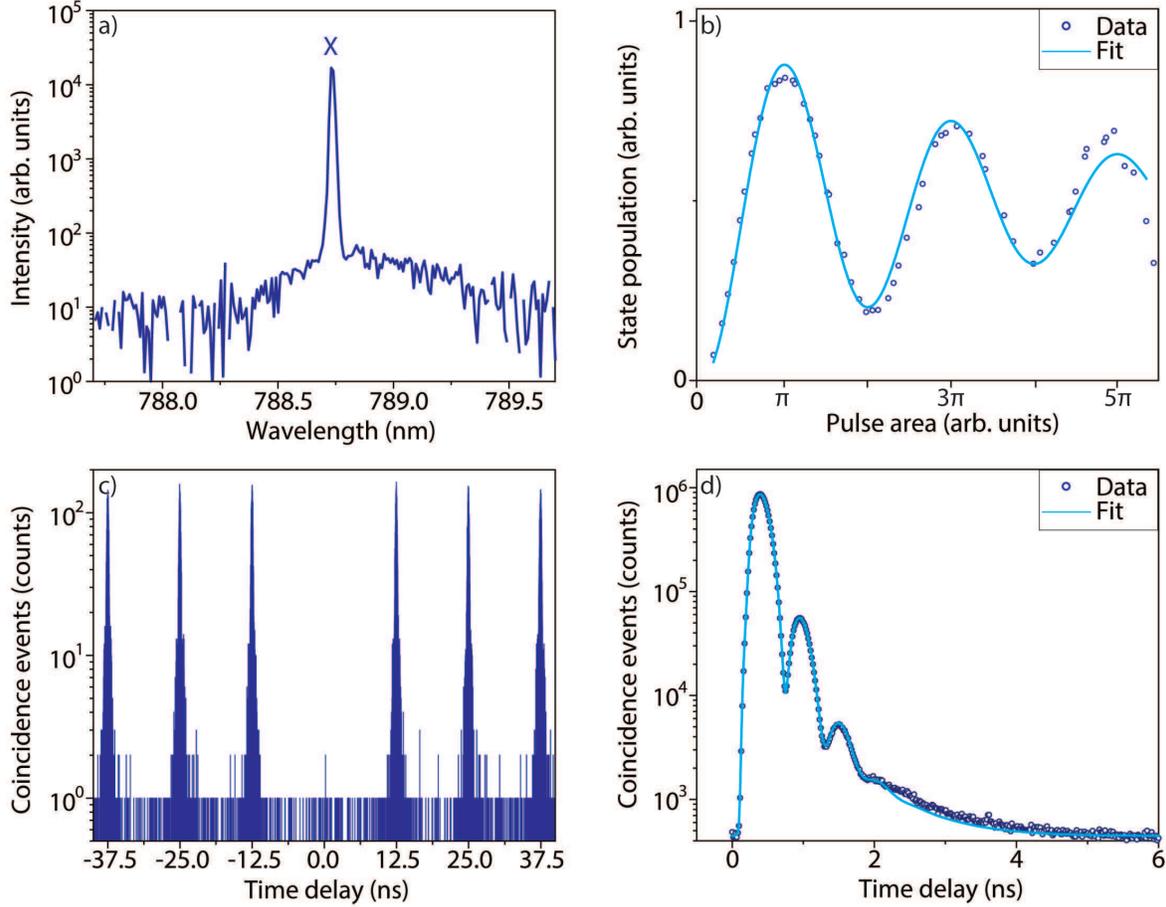}
    \caption{Characterization of the neutral exciton under pulsed resonant s--shell excitation. a) Resonance fluorescence spectrum in a semi logarithmic plot. b) Excitation laser power--dependent Rabi--oscillation up to a pulse area of 5$\pi$. From our fit we extract an occupation probability of 85\,\si{\percent} under $\pi$--pulse excitation.
    c) Second--order intensity correlation histogram yielding $g^{(2)}(0)=\left(1.8\pm0.18\right)\times10^{-3}$ d) Semi--logarithmic plot of the lifetime measurement with oscillations due to the fine structure splitting. The fit gives a lifetime of $196\pm2$\,\si{\ps} and a fine structure splitting of $7.44\pm0.05$\,\si{\micro\eV}.
    }
    \label{fig:fig2}
\end{figure}

In Fig.\,\ref{fig:figure1}~c) we show the spectrum of QD1 subject to pulsed non--resonant excitation at a wavelength of $781.\,\si{\nm}$. The neutral exciton (X), emitting at $788.73\,\si{\nm}$, is isolated from the rest of emission lines at longer wavelengths, which are attributed to electron-hole recombination in presence of additional carriers in the quantum dot which stem from nearby confined states due to the weak confinement in shallow quantum dots~\cite{Musial.Gold.ea:2014}. The trion (T) transition corresponds to the peak at $790.02\,\si{\nm}$.
To resonantly excite a s--shell transition (X or T) we tune the energy of the excitation laser to the transition´s energy, ideally resulting in photoluminescence only from this specific transition. 
Furthermore, the dephasing due to interactions with phonons and nearby trapped charge carriers is reduced~\cite{Ates.Ulrich.ea:2009a}. To stabilize the electric environment of the dot, we additionally illuminate the sample with a low intensity white light source~\cite{Gazzano.MichaelisdeVasconcellos.ea:2013, Reindl.Jons.ea:2017}. This results in a very clean spectrum with only one prominent line plotted semi logarithmically in Fig.\,\ref{fig:fig2}~a). 
To show that this excitation scheme addresses the quantum dot coherently, we perform power dependent pulsed resonance fluorescence measurements. In Fig.\,\ref{fig:fig2}~b) Rabi oscillations of the integrated intensity of the neutral exciton transition as a function of the excitation pulse area are shown. By exciting the quantum dot with a power corresponding to a pulse area of $\pi$, the population of the two--level system of the quantum dot is maximally inverted, preparing the quantum dot in the excited state. The procedure used to fit the data is explained in the supplementary. 
We extract a population probability for the neutral exciton state of \SI{85}{\percent} under $\pi$--pulse excitation. For all further measurements, the quantum dot is excited with a $\pi$--pulse.

%Pure single Photons: X g2
The second--order intensity correlation function shown in Fig.\,\ref{fig:fig2}~c) shows nearly background free single photon emission. By dividing the area of the center peak by the average area of eight side peaks, a measured degree of second--order coherence of $g^{(2)}(0)=\left(1.8\pm0.18\right)\times10^{-3}$ is determined. 
We attribute the deviation from $0$ solely to re-excitation~\cite{Fischer.Muller.ea:2016} and conclude that there is negligible residual excitation laser present in the correlation measurement.

%Lifetime X
Figure\,\ref{fig:fig2}~d) shows the lifetime measurement of the resonantly excited neutral exciton in a semi logarithmic plot. The exponential decay is observed with a periodic modulation~\cite{Dada.Santana.ea:2017}.
Due to the exchange interaction between the electron and hole spins, the degeneracy of the exciton states of the quantum dot is lifted leading to two linearly cross--polarized fine structure (FS) states with an energy splitting of $E_\text{FSS}$. 
During excitation, the spin of the exciton state is determined by the polarization of the excitation pulse~\cite{Kodriano.Schwartz.ea:2012,Muller.Kaldewey.ea:2013}. Then, the spin starts precessing on the equator of the Bloch sphere, oscillating between the two orthogonal fine structure states. The polarization of the emitted photon is set by the spin at the time of the recombination. Since one polarization component is suppressed by the polarizers in our setup, the intensity of the detected signal oscillates with a frequency $\nu=\frac{E_\text{FSS}}{h}$~\cite{Flissikowski.Hundt.ea:2001}.
The data is modelled with a fit explained in the supplementary, which yields a lifetime of $196\pm2$\,\si{\ps} and a fine structure splitting of $7.44\pm0.05$\,\si{\micro\eV}.\\

\begin{figure}[htbp]
    \centering
    \includegraphics[width=170mm]{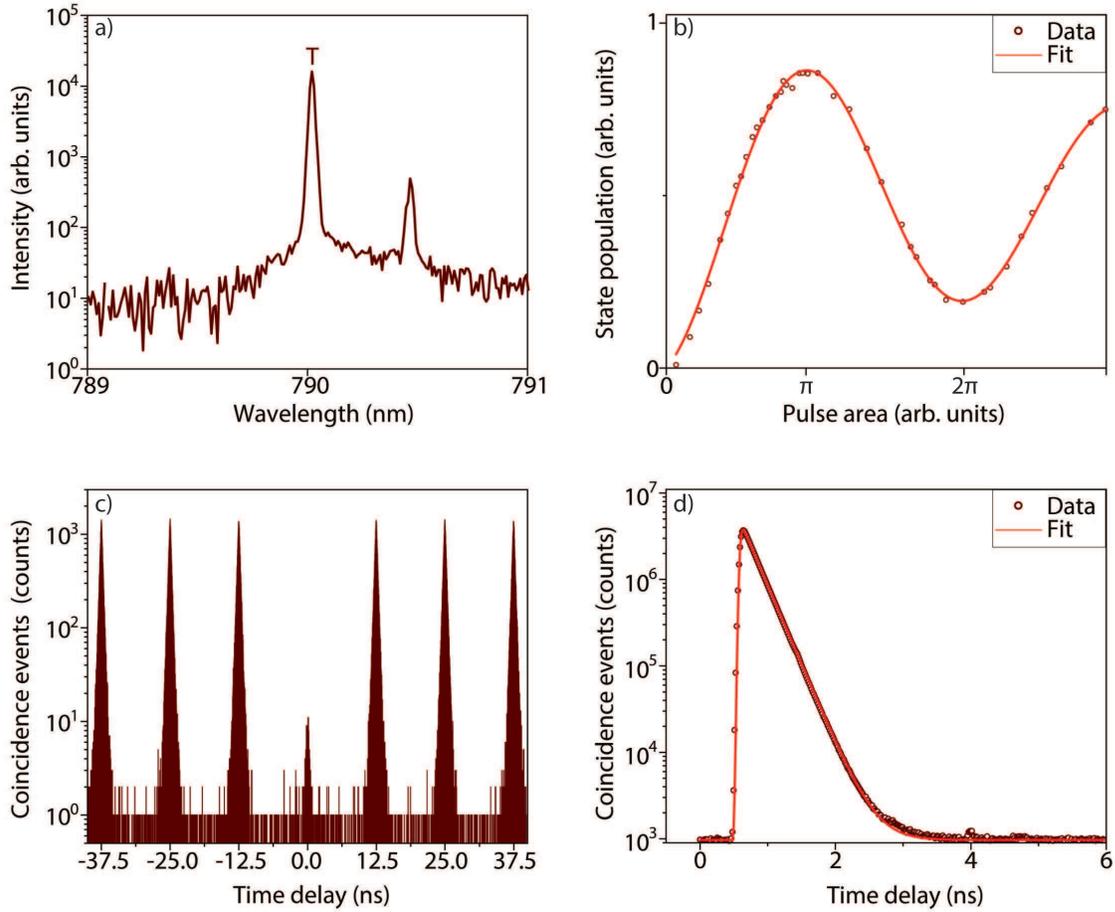}
    \caption{Characterization of the trion under pulsed s--shell resonant excitation. a) Resonance fluorescence spectrum in a semilogarithmic plot. The origin of the $\approx 30\times$ less intense line is discussed in the main text. b) Rabi--oscillation up to a pulse area of 3$\pi$. c) Second--order intensity correlation histogram yielding $g^{(2)}(0)=\left(6.4\pm0.38\right)\times10^{-3}$ d) Semi logarithmic plot of the lifetime measurement fitted with a single exponential decay giving us a lifetime of $236\pm2\,\si{\ps}$.
    }
    \label{fig:fig3}
\end{figure}

%Trion
In order to point out similarities with and differences to the neutral exciton, we investigate the resonantly excited charged exciton as well. 
Figure\,\ref{fig:fig3}~a) shows the spectrum of the trion of QD1 under pulsed resonant excitation. We observe an additional line close to the trion transition with $\approx 30$ times lower intensity, which might be a higher charge state emitting after a second capture process.
Similar to the neutral exciton we observe clear Rabi oscillations as a function of the excitation pulse area, as shown in Fig.\,\ref{fig:fig3}~b), and a maximum population inversion probability of $86\,\si{\percent}$ under $\pi$--pulse excitation. 
The second--order intensity correlation function yields a measured degree of second--order coherence of $\left(6.4\pm0.38\right)\times10^{-3}$ as shown in Fig.\,\ref{fig:fig3}~c) conforming that the laser suppression is very good.
In Fig.\,\ref{fig:fig3}~d) we show a semi logarithmic plot of a lifetime measurement. A single exponential fit gives a lifetime of $236\pm2\,\si{\pico\second}$.
As opposed to the lifetime measurement of the neutral exciton in Fig.\,\ref{fig:fig3}~d), this measurement shows no quantum beats, due to the lack of fine structure splitting of the trion state~\cite{Bayer.Ortner.ea:2002}.

\begin{figure}[htbp]
    \centering
    \includegraphics[width=170mm]{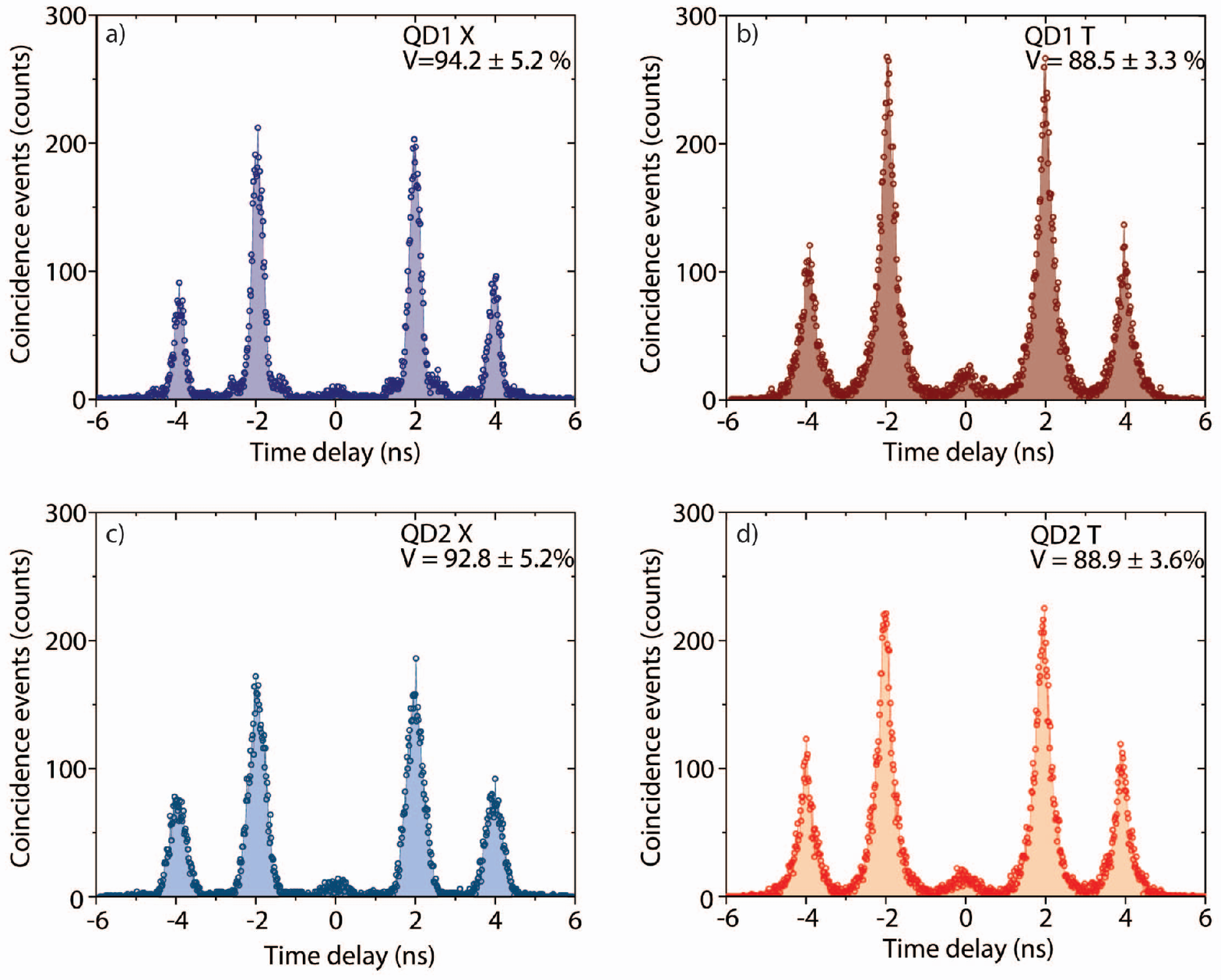}
    \caption{Hong--Ou--Mandel measurements under resonant s--shell excitation for the a) neutral exciton of QD1 $(V=94.2\pm 5.2\, \si{\percent})$,  b) trion of QD1 $(V=88.5\pm 3.3\, \si{\percent})$, c) neutral exciton of QD2 $(V=92.8\pm 5.2\, \si{\percent})$ and d) trion of QD2 $(V=88.9\pm 3.6\, \si{\percent})$.}
    \label{fig:fig4}
\end{figure}

%INDISTINGUISHABILITY
Having confirmed low multi--photon emission probability for both neutral and charged excitons under $\pi$--pulse resonant excitation, we continue to investigate the indistinguishability of consecutively emitted photons using a two--photon interference measurement as described above.
In Fig.\,\ref{fig:fig4}~a) and b) the center peak quintuplet for two--photon interference measurements of the neutral exciton and trion of QD1 are shown. The relative peak heights originate from different combinations of long and short paths in the Mach--Zehnder interferometer two consecutive photons can take. 
In the Hong--Ou--Mandel measurement of the neutral exciton, the same oscillations as in the lifetime measurements are visible. 
%The shape of the peaks in the Hong--Ou--Mandel measurements stems from the radiative decay of the investigated transition, which in turns mean that for the measurement of the neutral exciton, the same oscillations as for the lifetime measurements are visible.
In the limit of $g^{(2)}(0)=0$, the visibility of two-photon interference can be calculated from the area of the three center peaks A$_{1,2,3}$ by $V=1-\frac{2\times A_2}{A_1+A_3}$, where $V=100\,\si{\percent}$ corresponds to perfectly indistinguishable photons~\cite{Santori.Fattal.ea:2002}.
The uncorrected visibilities for QD1 are $V_\text{raw}=94.2\,\si{\percent}$ with a statistical error $\pm 5.2\,\si{\percent}$ for the neutral exciton and $V_\text{raw}=(88.5\pm 3.3)\,\si{\percent}$ for the trion. To compensate for imperfections in the setup, we measure the classical interference fringe visibility with a narrow continuous wave diode laser to be $(97.5 \pm 0.1)$\,\si{\percent} which yields the upper measurable bound of the visibility in this setup. 
We find additional evidence that the trion photons exhibit a lower indistinguishability by performing the Hong--Ou--Mandel measurements on a second dot as shown in Fig.\,\ref{fig:fig4}~c) and d). Here we obtain a visibility of $V_\text{raw}=(92.8\pm 5.2)\,\si{\percent}$ for the neutral exciton and $V_\text{raw}=(88.9\pm 3.6)\,\si{\percent}$ for the trion. One possible explanation of the lower indistinguishability of the trion photons in our Hong--Ou--Mandel measurements of consecutive photons with short time delays might be the longer lifetime of the trion state compared to the exciton state.
We would like to note that fitting the Hong--Ou--Mandel data, instead of summing up the data in specified time windows, very often overestimates the two-photon interference visibility, in particular if the data was collected with low timing resolution. Especially, when low time resolution is masking quantum beats and the dip at zero time delay, fitting can wrongly increase the visibility and even lead to unphysical results, i.e. visibilities above $100\,\si{\percent}$. This result is even independent of the used fit function (see details in the supplemental material).

%SUMMARY OUTLOOK CONCLUSION
We point out that our two--photon interference visibility value of $V=(94.2\pm 5.2)\,\si{\percent}$ is the highest raw value measured for any on--demand source without a micro--cavity and marks an important milestone for quantum dots derived from droplet etching. Near--unity indistinguishability was the last missing quantum optical property to put GaAs quantum dots on the horizon for future quantum communication and quantum information processing applications. Based on our results we foresee that other photonic structures than cavities, e.g. waveguides, trumpets, and nanowires~\cite{Gregersen.McCutcheon.ea:2016,Bulgarini.Reimer.ea:2014} to enhance light extraction efficiency from solid--state emitters, can be used to achieve even higher levels of indistinguishability without the need of Purcell enhancement. 

\section*{Author Information}
\subsection*{Corresponding Author}
Klaus D. J\"ons, e-mail: klausj@kth.se

\subsection*{Author contributions}
E.S. and K.D.J. built the setup with the help from K.D.Z., L.S, and T.L. E.S., L.H., L.S., K.D.Z, and K.D.J. performed the measurements. E.S., L.H., L.S, K.D.Z, carried out the data analysis with the help from K.D.J. S.F.C.S. and A.R. designed, optimized, and grew the quantum dot sample. M.R. and R.T. characterized the quantum dot sample, helping to optimize the sample. T.L. fabricated the final device with the help from R.T. E.S and K.D.J. wrote the manuscript with help from all the authors. K.D.J. conceived the experiment and supervised the project.

\subsection*{Notes}
The authors declare no competing financial interests.

\begin{acknowledgement}

This project has received founding from the European Union's Horizon 2020 research and innovation program under grant agreement No. 820423 (S2QUIP), the European Research Council (ERC) under the European Union’s Horizon 2020 Research and Innovation Programme (SPQRel, grant agreement no. 679183), the FWF (P 29603, P 30459), the Linz Institute of Technology, the German Federal Ministry of Education and Research via the funding program Photonics Research Germany (contract number 13N14846), Q.Com (Project No. 16KIS0110) and Q.Link.X, the DFG via the Nanosystem Initiative Munich, the MCQST,  the  Knut  and  Alice Wallenberg  Foundation  grant  ”Quantum  Sensors”, the Swedish Research Council (VR) through the VR grant for international recruitment of leading researchers (ref: 2013-7152), and Linn\ae{}us Excellence Center ADOPT.
K.D.Z. gratefully acknowledges funding by the Dr. Isolde Dietrich Foundation. K.M. acknowledges support from the Bavarian Academy of Sciences and Humanities.
A.R. acknowledges fruitful discussions with Y. Huo, G. Weihs, R. Keil and S. Portalupi.

\end{acknowledgement}

%\bibliography{achemso-demo}

\bibliographystyle{achemso}
\bibliography{Bib_RF_GaAs_QDs_HOM.bib}

\end{document}